\documentclass[%
 reprint,
superscriptaddress,
showpacs,preprintnumbers,
nofootinbib,
 amsmath,amssymb, 
 aps,
 prd,
 longbibliography,
]{revtex4-1}

\usepackage{cancel}
\usepackage{accents}
\usepackage{mciteplus,slashed}
\usepackage{amssymb,cancel,amsmath}
\usepackage{dcolumn}% Align table columns on decimal point
\usepackage{bm}% bold math
\usepackage[caption=false]{subfig}
\usepackage{appendix}
\usepackage{physics}
\usepackage{booktabs}
\usepackage{feynmp-auto}
\usepackage[T1]{fontenc}	
\usepackage{csvsimple}
\usepackage[colorlinks = true,
            linkcolor = blue,
            urlcolor  = blue,
            citecolor = blue,
            anchorcolor = blue]{hyperref}
\usepackage[section]{placeins}
\usepackage[capitalise]{cleveref}
\usepackage{booktabs}
\usepackage{graphicx}
\usepackage{mathrsfs}
\usepackage{comment}
\graphicspath{{Figures/}}

\AtBeginDocument{
\heavyrulewidth=.08em
\lightrulewidth=.05em
\cmidrulewidth=.03em
\belowrulesep=.65ex
\belowbottomsep=0pt
\aboverulesep=.4ex
\abovetopsep=0pt
\cmidrulesep=\doublerulesep
\cmidrulekern=.5em
\defaultaddspace=.5em
}
\usepackage[dvipsnames]{xcolor}
\usepackage[normalem]{ulem}
\usepackage{fontawesome} % for code link icons                                                

\def\iu{\mathrm{i}}

\newcommand{\newtext}[1]{\textcolor{black}{{\bf } #1}}

\begin{document}

\title{Empirical capture cross sections for cosmic neutrino detection with $^{151}{\rm \bf Sm}$  and  $^{171}{\rm \bf Tm}$ }
%\thanks{A footnote to the article title}%

\author{Vedran Brdar}
\email{vbrdar@fnal.gov}
\affiliation{Department of Physics and Astronomy, Northwestern University, 2145 Sheridan Road, Evanston, IL 60208, USA}
\affiliation{Theoretical Physics Department, Fermilab, Batavia, IL 60510,USA}

\author{Ryan Plestid}
\email{rpl225@g.uky.edu}
\affiliation{Department of Physics and Astronomy, University of Kentucky  Lexington, KY 40506, USA}
\affiliation{Theoretical Physics Department, Fermilab, Batavia, IL 60510,USA}

\author{Noemi Rocco}
\email{nrocco@fnal.gov}
\affiliation{Theoretical Physics Department, Fermilab, Batavia, IL 60510,USA}

\preprint{FERMILAB-PUB-22-026-T}
\preprint{NUHEP-TH/22-01}

\begin{abstract}
The nuclei $^{151}$Sm and $^{171}$Tm have been identified as attractive candidates for the detection of the cosmic neutrino background. Both isotopes undergo first-forbidden non-unique beta decays which inhibits a prediction of their spectral shape using symmetries alone and this has, so far, obstructed a definitive prediction of their neutrino capture cross sections. In this work we point out that for both elements the so-called ``$\xi$-approximation'' is applicable and this effectively reduces the spectral shape to deviate by at most $1\%$ from the one that would arise if beta decays were of the allowed type. Using measured half-lives we extract the relevant nuclear matrix element and predict the neutrino capture cross sections for both isotopes at $1\%$ level, accounting for a number of relevant effects including radiative corrections and the finite size of the nuclei. We obtained 
$(1.12\pm 0.01)\times 10^{-46}{\rm cm}^2$ for $^{171}$Tm and $(4.77\pm 0.01)\times 10^{-48}{\rm cm}^2$ for $^{151}$Sm. This method is robust as it does not rely on the data points near the end-point of the beta spectrum which may be contaminated by atomic physics effects, namely shake-up and shake-off. Finally, we calculate the target mass which is necessary for cosmic neutrino discovery and discuss several bottlenecks and respective solutions associated to the experimental program. We conclude that the detection of cosmic neutrino background by neutrino capture on $^{151}$Sm and $^{171}$Tm is achievable and free from theoretical limitations but still subject to technical issues that should be further investigated by the experimentalists in the context of the proposed PTOLEMY project.
\end{abstract}

%\pacs{Enter valid pacs numbers}% PACS, the Physics and Astronomy
                             % Classification Scheme.
%\keywords{Suggested keywords}%Use showkeys class option if keyword
                     	         %display desired
 \maketitle 
 %\section{General points to include}
 %
 %\begin{itemize}
 %   \item 
 %   \item 
 %\end{itemize} 
 
 \section{Introduction}
 \label{sec:intro}
The cosmic neutrino background (C$\nu$B) is a long sought after relic of the early universe. A number of theoretical proposals have been put forward, see for instance 
\cite{Gelmini:2004hg,Ringwald} (and for more recently proposed techniques \cite{Akhmedov:2019oxm,Bauer:2021uyj}), among which the method first discussed in 
\cite{Weinberg} that was further refined in \cite{Cocco:2007za} stands out.
This conventional detection scheme relies on neutrino capture on a long-lived, but unstable, beta emitter whose neutrino capture cross section is sizeable. The signature is the detection of electron/positron lying $2 m_\nu$ above the end-point of the beta spectrum. The primary considerations for C$\nu$B detection are $(i)$ an unstable, but long-lived isotope, $(ii)$ a detection scheme with a high energy resolution $\Delta E\lesssim 0.05$ eV, and $(iii)$ a large neutrino capture cross section and/or the ability to produce the desired isotope in large quantities. The two requirements listed under $(iii)$ are degenerate since a larger capture cross section allows for a smaller target mass for a fixed C$\nu$B detection yield. Other desirable attributes are a stable daughter nucleus and advantageous chemical properties for e.g.\ binding onto a substrate.

While not strictly required for C$\nu$B detection, a small $Q$ value is helpful primarily because beta decay lifetimes generically increase as $Q$ tends to smaller values. Moreover, a small $Q$ value results in a lower electron kinetic energy near the endpoint, for which it is easier to obtain the desired energy resolution $\Delta E\lesssim 0.05$ eV. Another desirable, but not strictly necessary, property for a ``good'' C$\nu$B target isotope is a relatively precise prediction for the neutrino capture cross section. Majorana and Dirac neutrinos differ in their capture rates by a factor of $2$ \cite{Long:2014zva}, and uncertainties of the neutrino overdensity in the vicinity of Earth attain $\sim 10\%$ precision \cite{deSalas:2017wtt}. A prediction for the capture cross section with a similar level of precision, i.e.\ $\sim 10\%$, would allow for the unambiguous determination of the Dirac vs. Majorana nature of neutrinos. 

The conventional choice for a target material is tritium due to its low $Q$ value, $Q(\!~^3H)=18.589 8(12)$ keV, long lifetime $t_{1/2}(\!~^3H) = 12.3$ yr, and calculable neutrino capture cross section $(\sigma v)_\nu = 38.34\times 10^{-46}\,\text{cm}^2$ \cite{Long:2014zva}. The PTOLEMY collaboration \cite{PTOLEMY:2018jst,PTOLEMY:2019hkd} expects a requirement of $100$ g of tritium, while only a few grams are currently suppliable in KATRIN's \cite{KATRIN:2019gru} windowless gaseous tritium source \cite{katrin}. The chief technical difficulty, however, is not tritium production but rather tritium packing \cite{Tully}. It is not guaranteed that all of the technical hurdles that face a tritium based detector will be overcome. It is of high importance, therefore, to keep a flexible perspective and to consider alternative isotopes for C$\nu$B detection. 

The requirements outlined above limit the potential list of viable isotopes to a relatively small subset of possibilities. Of these, the PTOLEMY collaboration is actively considering $^{171}$Tm \cite{Tully}, which can be efficiently produced by irradiating enriched erbium \cite{report}; alternative production methods are discussed in \cite{book2} and it should be stressed that patents exist for semi-industrial scale production conceived for nuclear power and medical applications \cite{patent}. In \cite{Cheipesh:2021fmg}, the authors advocate for heavy elements, namely $^{171}$Tm and $^{151}$Sm, due to spectral smearing arising from the zero point motion (ZPM) from low-energy intramolecular bonding.  It is interesting to note that both materials are metallic, and that $^{151}$Sm is a common byproduct of spent nuclear fuel and so supply issues are non-existent. Whether ZPM turns out to be a serious hurdle for C$\nu$B detection remaints to be seen, however the idea of using heavy nuclei, as opposed to tritium, does offer increased design flexibility for the experimental collaborations. 

What is lacking in the literature, however, is a precise prediction for the neutrino capture cross sections on heavy nuclei. In this work we supply a sub-1\% level determination of the neutrino capture cross sections for $^{151}$Sm and  $^{171}$Tm and compare them to the predictions for tritium. Our considerations apply to any heavy nucleus with a small $Q$-value satisfying $Q\ll Z\alpha/R$ where $Q$ is the total amount of kinetic energy released in the beta decay \cite{Mougeot}, $\alpha$ is the fine structure constant, $Z$ is the atomic number, and $R = (1.2 ~{\rm fm}) \times A^{1/3}$ is a typical nuclear radius. This opens the possibility of using heavy nuclei for C$\nu$B detection and reduces the issue of isotope selection to purely practical considerations e.g.\ the ability to produce sufficiently large yields of the given isotope. 

The beta decays of $^{171}{\rm Tm}$ and $^{151}{\rm Sm}$ are both first-forbidden non-unique since both transitions \newtext{preserve spin while flipping parity, i.e.\ $^{171}{\rm Tm}\rightarrow ^{171}{\rm Yb}$ is  $1/2+\rightarrow 1/2-$ and $^{151}{\rm Sm}\rightarrow ^{171}{\rm Sm}$ is  $5/2+\rightarrow 5/2-$ (see e.g.\ \cite{Behrens} for a textbook discussion)}. As was noted in \cite{Mikulenko:2021ydo}, this means that symmetry arguments are not sufficient to determine the spectral shapes. For heavy nuclei with small $Q$-values, however, the spectrum of the beta decays has an allowed shape up to corrections of $O(1/\xi)$ with $\xi = Z\alpha/(Q R)$ \cite{Mougeot,Behrens}. Corrections can be larger than what would be naively expected based on formal power counting, however this occurs only due to accidental cancellations in the leading order amplitude \cite{Kotani:1959zz,Behrens}. A broad survey of nuclei demonstrates that the ``$\xi$-approximation'' is reliable qualitatively across a broad range of heavy nuclei (many with $Q\gtrsim 300$ keV) \cite{Mougeot} for which $\xi\gtrsim O(10)$, and is quantitatively accurate at the level of a few percent for nuclei with $\xi\gtrsim O(100)$;  both $^{171}{\rm Tm}$ and $^{151}{\rm Sm}$ satisify this latter more stringent constraint. We discuss this as well as further effects that present a relevant correction to the beta decay spectrum in \cref{sec:shape}. In \cref{sec:data} we discuss previous measurements of $^{171}{\rm Tm}$ and $^{151}{\rm Sm}$ beta decays and show that they support of  the $\xi$-approximation up to systematic experimental uncertainties. \Cref{sec:xsec} is chiefly dedicated to a prediction of the neutrino capture cross sections for $^{171}{\rm Tm}$ and $^{151}{\rm Sm}$. These serve as a case study for any heavy nuclei with first forbidden non-unique decays that may be considered in the future. We include all relevant shape corrections at the level of $1\%$, and quantify uncertainties for the C$\nu$B  capture cross section accounting for relevant atomic physics effects such as shake-up and shake-off (\cref{subsec:shake}) which can transfer a portion of the C$\nu$B capture events below the natural beta decay endpoint. Armed with the cross section, in \cref{subsec:detection} we calculate required $^{171}{\rm Tm}$ and $^{151}{\rm Sm}$ detector mass and outline experimental techniques that could be employed for the successful C$\nu$B discovery. Finally, in \cref{sec:conclusions} we summarize our findings.

\section{Beta Decay Shape} 
\label{sec:shape}
The neutrino capture cross section and beta decay matrix elements are related by crossing symmetry. Hence, extracting the matrix element from beta decay measurement allows for the prediction of the relevant C$\nu$B detection cross section. Because all spectral distortions result from the electron final-state kinematics, they are common between $\nu$-capture and beta decay. We may therefore write the differential decay rate, $\dd \Gamma$, as 
\begin{equation}
    \label{crossing}
    \dv{\Gamma}{W_e}= \frac{E_\nu p_\nu}{\pi^2} \times (\sigma v)_\nu \times G(W_e)~,
\end{equation}
where $W_e = m_e + T_e$ with $T_e$ representing electron kinetic energy,  $(\sigma v)_\nu$  is the neutrino capture cross section at threshold i.e.\ for $\vb{p}_\nu =0$,  $E_{\nu}$ and $p_\nu$ are the the neutrino energy and three-momentum, respectively. The beta decay end-point energy is denoted by $W_e^{\rm max}$ and  $G(W_e)$ is a function that satisfies $G(W_e^{\rm max}) = 1$ and is otherwise determined by details of the matrix element governing the beta decay (i.e.\ it contains nuclear struture, Fermi function enhancements etc.). If $G(W_e)$ is known, then the neutrino capture cross section can be extracted from the beta decay half-life, $t_{1/2}$, since  
\begin{equation}
    t_{1/2} = \ln2\bigg/\int_{m_e}^{W_e^{\rm max}}\dd W_e  \dv{\Gamma}{W_e} ~. 
\end{equation}
In general, $G(W_e)$ contains both universally calculable functions such as the Fermi function, outer radiative corrections etc.\ and a nucleus-specific matrix element.

One can work in the long-wavelength (or equivalently point-like nuclear) limit which allows the full matrix element, $\iu \mathcal{M}=\!~_{\rm out}\hspace{-2pt}\braket{e A'}{A \nu}_{\rm in}$ to be reduced to a small number of nuclear matrix elements. If the transition is first-forbidden non-unique, and one neglects the effects of the nuclear Coulomb field, then there are six independent nuclear matrix elements that must be calculated each with different energy dependent prefactors \cite{Behrens}. This makes it impossible to predict the shape of $G(W_e)$ from first principles without further theoretical input. However, in practice, the nuclear Coulomb field \emph{cannot} be neglected as it dramatically alters the predicted spectral shape in addition to the well known Sommerfeld/Fermi function. This has been understood since at least the 1940s \cite{Konopinski:1943gum,Kotani:1959zz}, but was rigorously formalized by Behrens and B\"uhring in terms of radial integrals over electron radial wavefunctions that solve the Dirac equation with an extended charge distribution \cite{Behrens:1971rbq,Behrens}. The Coulomb field has two effects. First, the wavefunctions have different amplitudes near $\vb{r}=0$ than their normalization at large distances which is captured in the Fermi function $F(Z,W)$ (including finite size corrections). Secondly, in addition to a modified amplitude, the spatial variations of the wavefunctions are substantially altered. This effect modifies the convolution with the nuclear current density and as a consequence the behavior of the matrix element $\mathcal{M}$ as a function of $W_e$. The Behrens-B\"uhring formalism expands the matrix elements in three small parameters, $W_e R$, $m_e R$, and $Z\alpha$ \cite{Behrens}, the latter of which is much larger than the former two. Consequently, the prefactors appearing in front of the nuclear matrix element can be taken as energy independent and the beta decay spectrum reduces to the calculable allowed spectrum. This is the basis of the $\xi$-approximation. Corrections appear at $O(T_e R/Z\alpha)$ and are largest near the end point, being of relative size $1/\xi = Q R/Z\alpha$. 

The $\xi$-approximation has recently received renewed scrutiny due to its wide application in the study of heavy-nuclei beta decays. As noted by the authors in \cite{PhysRevC.100.054323} in the context of the reactor anomaly (see also \cite{Patrick}), \emph{a priori} it is only expected to be valid if $\xi \gg 1$, and for certain applications (such as the uranium decay chain) this condition is not satisfied \cite{PhysRevC.100.054323}. The author of \cite{Mougeot} conducted a systematic investigation of 53 nuclei, ranging from light- to heavy-elements with $Q$ values between 20 keV to 1.3 MeV. The conclusion was that the $\xi$ approximation generically holds at the expected level (i.e.\ up to $1/\xi$ corrections) and that many nuclei have $\xi \lesssim 25$ and hence feature 1\%-10\% level deviations from the allowed shape. In \cite{Mougeot} this is presented as a ``failure'' at the perecent level, however for C$\nu$B detection even a 10\% uncertainty on the cross section is likely suffcient for practical purposes. We therefore interpret the results of \cite{Mougeot} as providing support for the application of the $\xi$-approximation to heavy nuclei in the context of C$\nu$B detection, and note that for low-$Q$ nuclei the approximation is expected to hold at the level of 1\% or better.

The elements of interest in this paper, $^{171}{\rm Tm}$ and $^{151}{\rm Sm}$, have two beta decay branches; the final state nuclei ($^{171}{\rm Yb}$ and $^{151}{\rm Eu}$) can be either in the ground or first excited state where the former occurs in $\sim 98-99\%$ of the decays; all higher excited states are energetically forbidden.  Both nuclei have first-forbidden non-unique tranistions in both the ground state (primary) and excited state (secondary) branches. The secondary branches are rare and only enter our discussion through an overall normalization of the half-life. Only the primary branch is important for C$\nu$B detection because capture onto the second beta-branch would lie beneath a gigantic background from the primary beta-branch's neutrino decay. Hence, when we discuss $Q$ values, we will always refer to the ground state decay; $Q$-values for $^{171}{\rm Tm}$ and $^{151}{\rm Sm}$ beta decay are $96.5$ and $76.7$ keV, respectively and can be computed to high accuracy from isotopic mass measurements (nuclear binding energies are known to $0.1$ eV level precision). As we emphasize above, both $\!~^{171}{\rm Tm}$ and $\!~^{151}{\rm Sm}$ have low $Q$ values relative to the sizeable Coulomb potentials such that their $\xi$ values are very 
large,  $\xi(\!~^{151}{\rm Sm}) = 181.90$ and $\xi(\!~^{171}{\rm Tm}) = 154.37$. Thus, \emph{a priori} we expect their decay spectrum to have the same shape as an allowed decay up to small  O($1\%$) corrections.
In \cref{sec:data} we further strengthen this theory driven expectation with empirical evidence that points towards the validity of the $\xi$ approximation.

The shape of the allowed beta spectrum has recently received considerable theoretical attention with predictons for the spectral shape expected to be accurate at the level of 0.01\% across the full kinematic range \cite{Hayen:2017pwg}. Furthermore, two notable codes have been developed \cite{BetaShape,Hayen:2018lhg}. For our purposes sub-percent corrections to the allowed shape are irrelevant, being sub-dominant to $O(1/\xi)$ corrections to the nuclear matrix elements. We therefore include effects which modify the spectrum at the  $\sim 1\%$ level as quantified by Table VII of \cite{Hayen:2017pwg}, which leads to
\begin{align}
    \dv{\Gamma}{W_e} =& \, F_0(Z,W_e)\, L_0(W_e)\, R(W_e)\,  X(W_e)\,  r(W_e)\,  \nonumber \\& \hspace{0.4\linewidth}  W_e\, p_e\, E_\nu\,p_\nu \, \times C_0~.
     \label{shape}
\end{align}
Here, $F_0$ is the traditional Fermi function, $L_0$ accounts for the finite size of the nucleus, $R$ includes outer radiative corrections captured by Sirlin's $g$-function \cite{Sirlin:1967zza}, $X$ is an atomic exchange correction,  and $r$ an atomic mismatch correction. The detailed theoretical description of each of these terms is given
in \cite{Hayen:2017pwg}. Numerically, we find that $X$ yields up to $7\%$ correction at lower energies and the effect ceases toward the endpoint. The atomic mismatch function, $r$, gives $1-2\%$ effects across the whole spectrum while $R$ modifies the spectrum at the level of $3\%$. Finally, $L_0$ corrects the shape by $2\%$ at low energies while the correction gets smaller as one moves toward larger energies. $C_0$ is a constant that depends on the nuclear matrix elements mentioned above. For a non-unique first forbidden decay, $C_0$ would be replaced by the so-called ``shape factor'' $C(E)$. For large values of $\xi$ we have that $C(E)= C_0 + O(1/\xi)$. Comparing \cref{crossing} and  \cref{shape} we can infer $G(W_e)$ which is calculable up $1/\xi$ suppressed corrections. If we define $G(W_e) = \mathcal{G}(W_e)/\mathcal{G}(W_e^{\rm max})$ then we have that 
\begin{equation}
    \mathcal{G}(W_e)= \,  F_0(Z,W_e)\, L_0(W_e)\, R(W_e)\,  X(W_e)\,  r(W_e)\, W_e\, p_e\, \label{fancy-G}. 
\end{equation}
For our numerical implementation we compute the spectrum using \verb|BetaShape| software \cite{BetaShape} which incorporates effects parametrized by $F_0$, $L_0$ and $R$. The functions $X$  and $r$ are subsequently added by hand following the analytic approach outlined in \cite{Hayen:2017pwg}.

\section{Empirical support for  \\the allowed spectrum \label{sec:data}}

%%%%%%%%%%%%%%%%%%%%%%%%%%%%%%%%%%%%%%%%%%%%%%%%%%%%%%%%%%%%%%%%%%%%%%%%%%%%%%%%%%%%%%%%%%%%%%%%%
 \begin{figure}[t]
	\centering
	\includegraphics[width=\linewidth]{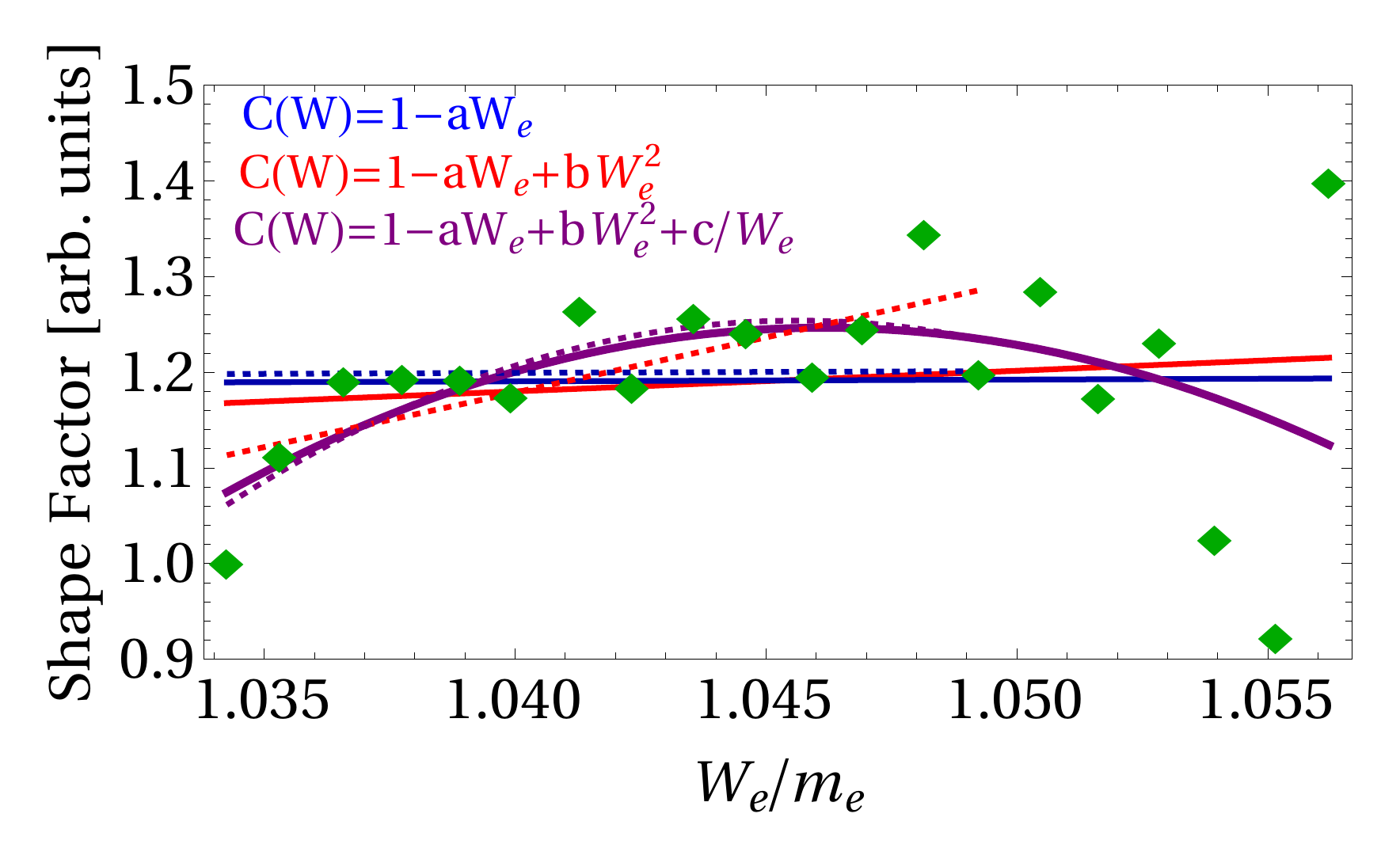} \\
        \includegraphics[width=\linewidth]{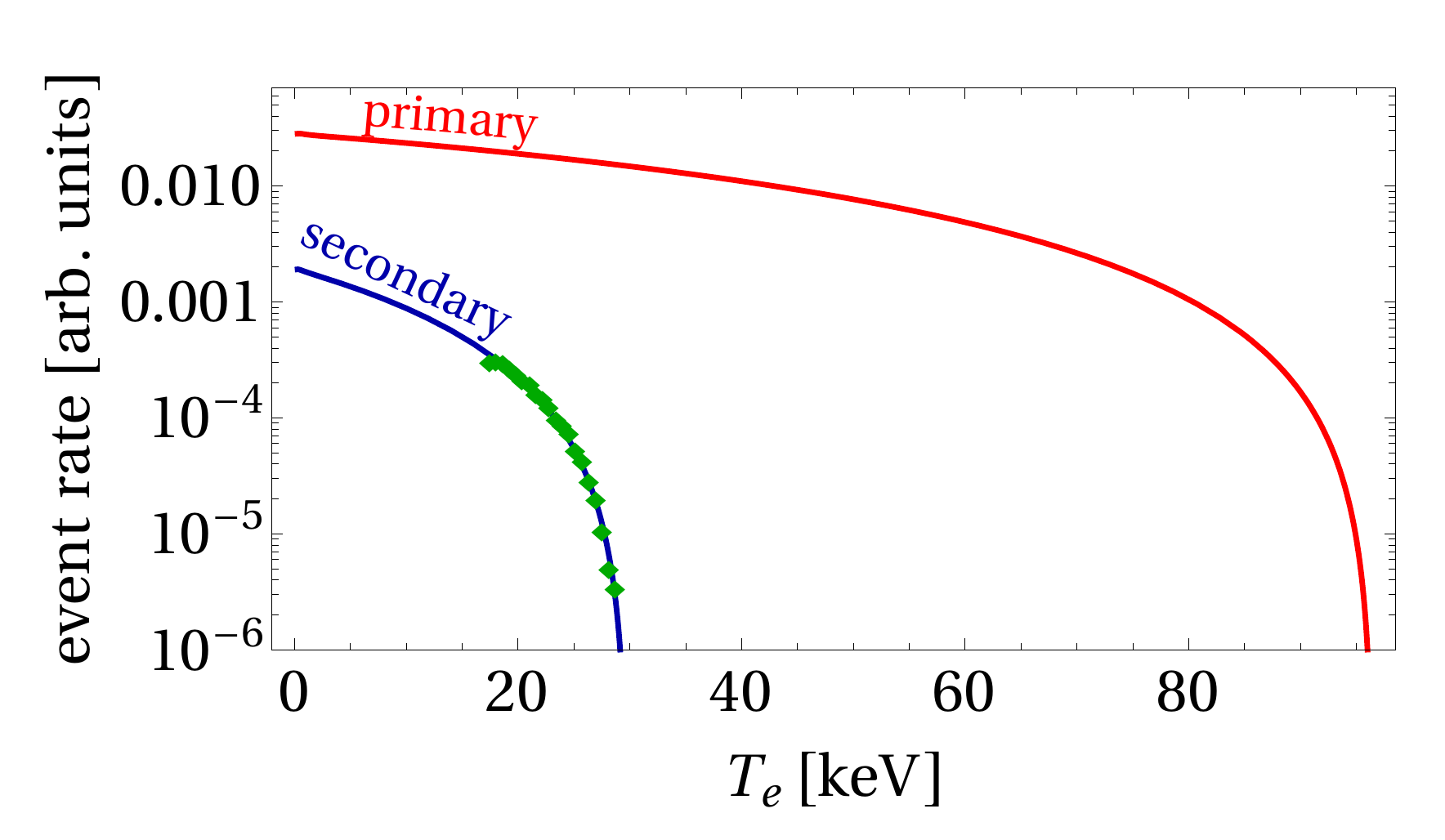}
        
	\caption{\textbf{Upper panel:} Data from \cite{Robinson} for the secondary beta branch of $^{171}$Tm i.e.\ the excited state decay (green diamonds) is show together with fits of several shape factor functions to data. Solid (dashed) curves represent the fit where all available data (only points further from the end-point energy) have been considered. \textbf{Lower panel:} Data superimposed atop the beta shape prediction for the primary and secondary beta branches of $^{171}$Tm. }
	\label{fig:1}
\end{figure}
 %%%%%%%%%%%%%%%%%%%%%%%%%%%%%%%%%%%%%%%%%%%%%%%%%%%%%%%%%%%%%%%%%%%%%%%%%%%%%%%%%%%%%%%%%%%%%%%%

In the previous section we have discussed formal power counting in the Behrens-B\"uhring formalism \cite{Behrens} which implies a theoretical error that scales parameterically as $1/\xi$. It is well known, however, that the $\xi$ approximation can fail due to approximate dynamical selection rules \cite{Behrens,Mougeot,Hayen:2018lhg}. To the best of our knowledge, there are no such dynamical selection rules for either of $^{151}$Sm or $^{171}$Tm. At a purely theoretical level, however, the possibility still remains that there is an accidental cancellation among amplitudes rendering the leading order terms in the $\xi$-approximation small, such that sub-leading corrections are larger than anticipated \cite{Kotani:1959zz}. It is therefore helpful to understand the agreement between theoretical expectations and published data for both nuclei. In this section we perform such a comparison and in both cases, the measured spectra are in agreement with the allowed spectrum once one accounts for experimental uncertanties, suggesting the applicability of the $\xi$-approximation. A similar empirical study could be performed \emph{in situ} in a future C$\nu$B detector by scanning across the beta spectrum. 

The beta decay of $^{171}$Tm was previously measured in \cite{Hansen,PhysRev.107.1314} where the authors chiefly focused on the decay into the excited state of $^{171}$Yb (secondary branch) where an electron is detected in association with the $x$-ray line from $^{171}$Yb deexcitation. While we were not able to find measured spectrum for the primary branch, the data for the secondary branch is available (Fig 17 in \cite{Robinson}) and it is a good proxy with which to test the $\xi$-approximation given that Q-values for both decays differ by only $\sim 60$ keV and therefore $\xi\gg 1$ in both cases. \newtext{We further note that the short lifetime of the excited state in $^{171}$Tm suggests large overlap between the ground state and excited state wavefunctions; this further supports using the secondary branch as a proxy for the primary branch.} By employing this data, shown in \cref{fig:1}, we can discuss compatibility of the measured shape with the allowed one. The data is appropriately corrected  by including effects beyond traditional Fermi function (see \cref{shape}) and we also rescale the energy scale by removing the energy of the X-ray (5.4 keV) that was emitted in the process. The authors of \cite{Robinson} corrected their spectrum to account for a systematic effect in their scintilation light yield.  The authors note that while their endpoint is in ``reasonable'' agreement (in fact it was 1.2 keV too large) with expectations from mass spectometry, that ``this agreement may be somewhat fortuitous since there were large resolutions corrections applied to the scintillation spectrometer data'' \cite{Robinson}. We therefore subtract 1.2 keV from the reported energy in \cite{Robinson} bringing the observed endpoint into agreement with expectations from mass spectrometry.

If the spectrum was of the allowed type one would expect to see data points forming a flat line with some scatter due to statistical fluctuations. The data in \cite{Robinson} does not include an overall normalization and therefore offers a ``shape only'' measurement with no quantified statistical or systematic uncertainties (as noted above systematic errors are siginficant with the experiment failing to properly capture the endpoint). Although we cannot precisely quantify the agreement with the allowed spectrum, we can bound the error from above. We observe at most $40\%$ deviations from such scenario which improves to the $\sim 10\%$ if one discards energies near the end point that necessarily suffer from larger statistical errors than the rest of the data. Although the number of the observed events per bin is not reported, nor is any measurement uncertainty, we may still perform fits using $3$ functions shown in  \cref{fig:1} and which are typically considered for the shape factor correction in the context of  first-forbidden non-unique beta decays \cite{Behrens,Mougeot,Morita}. This is closer to a qualitative than quantitative excercise because of our inability to assign meaningful numerical uncertainties to the data. We perform two fits: 1) taking into account all data (solid curves), and 2) fitting only the portion of the data further from the endpoint (dashed curves); here we omitted points that correspond to energies that are less than 3.5 keV from the endpoint. We show best fit lines only, however none of the parameterizations provide a markedly better fit than a flat line (the allowed approximation). In the absence of better data, we interpret this as supporting the validity of the $\xi$-approximation {\em at least} at the level of 10\% for the secondary beta branch of $^{171}$Tm.  Updated measurement, preferably for the ground state decay, could provide a useful cross check of this expected behavior and could potentially serve as an early nuclear physics target for a PTOLEMY-like demonstrator. 

The half-life of $^{151}$Sm was recently measured with high precision and the spectrum was compared to the allowed one \cite{Sm}. Upon including relevant atomic effects such as $X(W_e)$ and $r(W_e)$ (see again \cref{shape}) the authors found that the measurement yields less than $0.2\%$ deviation from the theoretical allowed shape across the whole spectrum (see Fig.\ 2 of \cite{Sm}).  As discussed in \cref{sec:shape}, this is precisely the order of magnitude one would expect from $1/\xi$ suppressed corrections to the allowed spectrum. We conclude, therefore, that empirical evidence \emph{strongly} supports theoretical expectations, and that the $\xi$-approximation can be safely applied to the beta decay of $^{151}$Sm.

\section{Cross Section Extraction \label{sec:xsec}}
\begin{figure}[t]
	\centering
        \includegraphics[width=\linewidth]{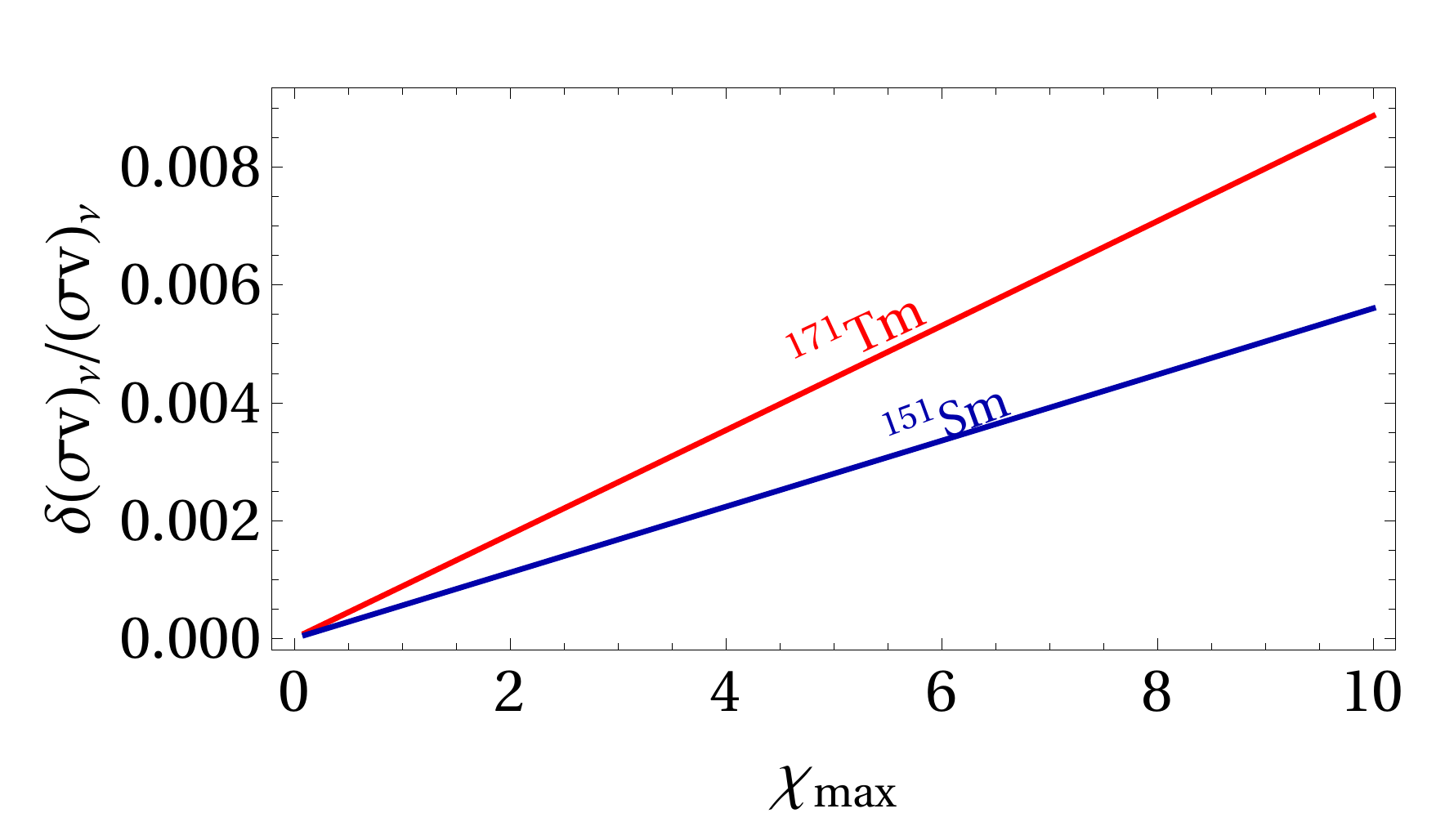}
        
	\caption{Relative error estimates for the neutrino capture cross section at threshold, $(\sigma v)_\nu$
          for both $^{151}$Sm and $^{171}$Tm as a function of the parameter $\chi_{\rm max}$. Our nominal choice in \cref{eq:Tm,eq:Sm} is $\chi_{\rm max}=4$. Even for very large values of $\chi_{\rm max}$ that would indicate an accidental cancellation of the leading order matrix element, we find relatively small errors. We conclude that the cross sections for both elements (and indeed most heavy nuclei) can be reliably extracted from the half life. }
	\label{fig2}
\end{figure}

The cross section for neutrino capture at threshold can be extracted from the measured half life of each isotope using \cref{shape,fancy-G,crossing}. Specifically, we consider the equation
\begin{align}
{\rm BR}_1 \,\frac{\ln 2}{t_{1/2}}=&\int_{m_e}^{W_{e}^{\rm max}} dW_e \bigg[ F_0(Z,W_e)\, L_0(W_e)\, R(W_e)\,  X(W_e)\, \nonumber \\&  r(W_e)\, W_e\, p_e\, E_\nu\,p_\nu \, \times C_0 (1 + \chi  W_e/(m_e \,\xi))\bigg]\,,
\label{eq:vv}
\end{align}
where ${\rm BR}_1$ is the branching ratio for the decay in the ground state, 
the square brackets contain corrections in \cref{shape}. Using \verb|BetaShape| and the additional analytic corrections discussed above we extract those functions specifically for the ground state decay (i.e.\ the primary branch). The term in brackets that depends on  $\chi$ parametrizes sub-dominant corrections to the $\xi$ approximation which may impact the cross section extraction. Here, $\chi$ is a dimensionless parameter that we vary in the range between $-\chi_{\rm max}$ and $+\chi_{\rm max}$; our nominal choice is $\chi_{\rm max}=4$, however we also discuss how the error estimate varies for different choices of $\chi_{\rm max}$ (see e.g.\ \cref{fig2}).  For each value of $\chi$, we can extract $C_0$ by solving \cref{eq:vv}. Demanding the observed half-life be reproduced introduces correlations between $C_0$ and $\chi$ that we take into account in our analysis. Performing a scan in both variables then gives us an array of tuples $(\chi,C_0)$ which are then used to predict $(\sigma v)_\nu$ via
\begin{align}
(\sigma v)_\nu =& \pi^2 \bigg[ F_0(Z,W_e) \, L_0(W_e)\, R(W_e)\,  X(W_e)\, r(W_e)\, \nonumber \\ &\, W_e\, p_e\times C_0 (1 + \chi  W_e/(m_e \,\xi)\bigg]\Bigg|_{W_e=W_e^{\rm max}}\,. 
\end{align}
Using this procedure we generate an ensemble of $(\sigma v)_\nu$ from which we extract the mean and standard deviation. Notice that the cross section is most sensitive to the high energy part of $\mathcal{G}(W_e)$ via the numerator. The denominator averages over all accessible electron energies and so effects which only alter the low-energy portion of the spectrum, e.g.\ atomic screening, still affect the cross section extraction, but only enter via an averaged quantity and so are subdominant. 

For $^{171}$Tm we use $t_{1/2} = 1.9216$ years,  ${\rm BR}_1 = 0.9804$, and $W_e^{\rm max} =96.5$ keV. For $^{151}$Sm we use $t_{1/2} = 88.8$ years, ${\rm BR}_1 = 0.9909$, and $W_e^{\rm max} =76.6$ keV. The values for ${\rm BR}_1$ and $W_e^{\rm max}$ are  adopted from \verb|BetaShape| whereas the half-lives quoted above are taken from nuclear data tables. 
 %These values have also their uncertainties  and we take them into account when reporting cross sections. 

Putting all of this together we find for  $^{171}$Tm, 
\begin{align}
    (\sigma v)_\nu &= (1.12\pm 0.01)\times 10^{-46}{\rm cm}^2~,  
    \label{eq:Tm}
\end{align}
while for $^{151}$Sm we have instead, 
\begin{align}
     (\sigma v)_\nu &=(4.77\pm 0.01)\times 10^{-48}{\rm cm}^2~\,.
     \label{eq:Sm}
\end{align}
Both predictions use our nominal choice of $\chi_{\rm max}=4$, while the relative error estimated for different choices of $\chi_{\rm max}$ is plotted in \cref{fig2}. We note that the uncertainty of the half-life and  ${\rm BR}_1$ as well as the fraction of events in the region associated to the uncertainty of the endpoint energy are all subdominant to errors stemming from $1/\xi$ corrections to the nuclear matrix element. Our analysis does not account for mismodelling of the allowed spectrum at low energies, or uncertainties in the $Q$ values. Nevertheless, these uncertainties are univeral among all beta emitters including those with allowed transitions; a detailed study is warranted if $^{151}$Sm or $^{171}$Tm are pursued further by experimental collaborations.

These estimates may be compared to the state of the art prediction for the tritium capture cross section of $38.34 \times 10^{-46}{\rm cm}^2$ \cite{Long:2014zva} which is clearly larger, but not drastically so. This already suggests that if $^{151}$Sm or $^{171}$Tm are employed as the target material, the required fiducial mass for the detection would exceed 100 grams that is the mass of tritium necessary for $O(10)$ events that would merit a discovery. The relevant merits of different detector materials are, however, more complicated than their bulk mass alone. In the next section we discuss effects relevant for the C$\nu$B discovery with $^{151}$Sm and $^{171}$Tm.

\section{Detection Prospects \label{sec:detector}}

%%%%%%%%%%%%%%%%%%%%%%%%%%%%%%%%%%%%%%%%%%%%%%%%%%%%%%%%%%%%%%%%%%%%%%%%%%%%%%%%%%%%%%%%%%%%%%%%%
 %%%%%%%%%%%%%%%%%%%%%%%%%%%%%%%%%%%%%%%%%%%%%%%%%%%%%%%%%%%%%%%%%%%%%%%%%%%%%%%%%%%%%%%%%%%%%%%%

In this section we consider both theoretical and practical issues related to the use of heavy nuclei. Recent work has suggested that zero point motion (ZPM) is a fundamental barrier to C$\nu$B detection with tritium and has used this as motivation for considering heavy nuclei \cite{Cheipesh:2021fmg}. We do not comment on ZPM motion here, and ignore binding to the substrate all together. Our discussion will instead focus on practical issues such as packing density, means of production, and low-lying atomic excitations that could deplete the C$\nu$B capture signal. 

\subsection{Detector material requirements \label{subsec:detection} }

A simple figure of merit for each nuclei is the mass or number of atoms required to detect 10 C$\nu$B neutrinos per year. In this section we assume, as is common practice, that the event yield is given by the neutrino capture cross section at threshold. In reality atomic shake-up and shake-off modify this picture which we discuss in more detail in \cref{subsec:shake}. Additional effects related to substrate binding such as smearing due to ZPM may also alter detector mass requirements but, as we have outlined above, we do not discuss these issues here. In addition to the raw detector mass, other practical considerations come into play. Chief among them, is the ability to produce the relevant nuclear isotope, and to achieve sample purity free of any other beta emitter whose end point lies above beta branch of interest. 

The simplest comparison one can make is to take the ratio of the neutrino capture cross sections at threshold $(\sigma v)_\nu$ for tritium and the heavy elements of interest, $^{171}$Tm and $^{151}$Sm; this was previously discussed in \cite{Mikulenko:2021ydo}. \newtext{The authors of \cite{Mikulenko:2021ydo} also obtained predictions for the capture cross sections on $^{171}$Tm and $^{151}$Sm (see footnote 7 of their paper) but argue that the capture cross sections on $^{171}$Tm and $^{151}$Sm cannot be predicted from first principles due to the transition being first forbidden non unique. As outlined above, we claim that reliable error estimates can be obtained by leveraging a systematic expansion of electron radial wavefunctions as introduced by Behrens and B\"uhring. We note that the same formalism underlies the typical power counting for allowed transitions where higher multipole operators are neglected. While it is always true that accidental fine tuning can upset formal power counting (in this case the series in $1/\xi$) for the nuclei we consider this would require a fine-tuning at the level of a few parts per thousand. The central values quoted in \cite{Mikulenko:2021ydo} agree with our results, whereas our main contribution is to place the extraction of a capture cross sections from a half life on firm theoretical footing for heavy nuclei and provide an estimate of its theoretical uncertainty.} The capture cross sections are quoted per nucleus, however the heavy nuclei are roughly 60 times as heavy as tritium and so the required detector \emph{mass} is enhanced by the same factor. Since the half-life of $^{171}$Tm is 1.92 yr we take into account its radioactive decay (these effects are relatively negligible for $^{151}$Sm whose half life is 88.8 yr). If we take the neutrino capture cross section from \cref{eq:Tm}, and account for the decay as a function of time we find
\begin{align}
    N_{^{171}Tm} = \frac{9.5\times 10^{26} \text{year} \,\,
    t_{1/2}^{-1}}{\text{Exp}[-t_{i}/t_{1/2}]-\text{Exp}[-t_{f}/t_{1/2}]}\,,
\label{eq:N}
\end{align}
where $t_{i}$ and $t_{f}$ denote start and end of the experiment's runtime, where $t=0$ corresponds to the production time of $^{171}$Tm which has significantly shorter half-life with respect to tritium. If we take $t_{i}=0$ and $t_{f}=1$ year, we find $1.2\times 10^{27}$ atoms corresponding to 350 kg of $^{171}$Tm to be required. For $^{151}$Sm, we can obtain the equivalent of \cref{eq:N} by employing \cref{eq:Sm} instead of \cref{eq:Tm} and we find that 6 tonnes of material ($N_{^{151}{\rm Sm}}=2.4\times 10^{28}$) is required. These numbers should be compared with 100 g or $2 \times 10^{25}$ atoms of tritium.

For both thulium and samarium it is clear from the above discussions that a much larger target mass will be needed to achieve comparable C$\nu$B detection as with tritium. One may reasonably wonder if the production of each isotope could serve as a bottleneck for the experiment. As we will now argue, both isotopes can easily be produced in the required quantities. The production of $^{171}$Tm was already proven successful in the 1960s  from irradiated enriched erbium \cite{report}. Further techniques have been discussed in \cite{book2} and a patent also exists \cite{patent} for relatively high purity production. However, all proposed $^{171}$Tm production mechanisms yield a roughly part per thousand contamination of $^{170}$Tm. This is a difficult problem because $^{170}$Tm cannot be removed by chemical extraction and purification is therefore difficult. Importantly $^{170}$Tm is also an unstable beta-emitter with a larger Q value ($314$ keV). This means that the signal region for C$\nu$B capture on the primary beta-branch of $^{171}$Tm will be swamped with background from the decay of $^{170}$Tm; we return to this point below. The isotope $^{151}$Sm has no such ``dangerous neighbors'', and advantageously appears as byproduct of the $^{235}$U decay chain, and so is present in all spent nuclear fuel. Left unprocessed, spent nuclear fuel is clearly ill-suited for C$\nu$B detection, however Sm can be extracted chemically. Research studies from Savanah River have found chemical techniques that can achieve contaminations as small as  $\sim 3 \times 10^{-4}$ while retaining a $\sim 50\%$ yield of $^{151}$Sm \cite{osti_774848}. Repeated reprocessing naively gains in purity multiplicatively, such that four cycles would yield a $\sim 10^{-14}$ level contamination, and a $\sim 12.5\%$ yield. Further studies are required to determine the ultimate capabilities of spent nuclear fuel purification if $^{151}$Sm is deemed an attractive candidate.

The fact that $^{170}$Tm is chemically indistinguishable from $^{171}$Tm suggests that purification may be difficult. A brute force solution would be to leverage the shorter lifetime of $^{170}$Tm ($t_{1/2}\sim 129$ days) in comparison with $^{171}$Tm. If one waits long enough after producing the target material, the fraction of 
$^{170}$Tm would eventually become small enough such that the detection of C$\nu$B via $^{171}$Tm would not be compromised. This waiting period could be done with a bulk sample, and the $^{170}$Yb and $^{171}$Yb that results could be subsequently removed with chemical methods. We estimate optimal signal to background in a period roughly $10-20$ years after production, and during this time substantial quantities of $^{171}$Tm would decay which would then push the required target mass upwards to masses on the order of a kilotonne. Taken together, these conditions may invalidate $^{171}$Tm as realistic C$\nu$B detection material for experimental applications, however improved purification techniques could modify this conclusion.

Overall, the required detector masses for the elements considered in this work do not exceed the magnitude of the present and near future neutrino detectors. Hence, an experimental realization in which either $^{171}$Tm or $^{151}$Sm  would be employed does not appear unrealistic on grounds of detector mass requriements alone. This is, however, not the primary limiting factor for a PTOLEMY-like experiment. C$\nu$B detection demands that the signal electron does not scatter while passing through target material. This requirement has driven experimental designs towards a modular design with packing on substrates. The spatial extent of these modules then defines the fiducial region in which the experiment must operate e.g.\ maintaining electric and magnetic fields. We now turn to this issue in the context of heavy nuclei. 

\subsection{Packing and layering} 

Current experimental designs from PTOLEMY rely on a highly efficient loading of tritium atoms onto a graphene substrate ($\geq 45\%$ graphene sites loaded \cite{ZHAO2021244}). Pieces of graphene can then be assembled into cells with embedded wires responsible for the transport of the signal electrons from the inter-cell vacuum, to regions with no detector material. This must be efficient to avoid signal electrons encountering detector material and losing energy; even a 0.5 eV energy loss would prevent the detection of a C$\nu$B signal electron. This effect is often termed ``backscatter'' and has been well studied for tritium targets in the Mainz Neutrino Mass Search collaboration \cite{Kraus:2004zw}.

For high-$Z$ elements, backscatter becomes a serious concern. The probability for an electron to scatter inelastically is given by $n_\perp \sigma_{\rm inel}$ where $n_\perp$ is the 2-D number density of scattering targets and where the inelastic cross section is given by $\sigma_{\rm inel}= \int \dd W  \dd \sigma_{\rm inel}/\dd W$ where $W$ is the energy transfer. This differs from the conventional input in the Bethe theory of ionization where the integrand is weighted by $W$. For C$\nu$B detection \emph{any} atomic excitation, no matter how small, will spoil the detection of a signal electron. Inelastic cross sections for $Z\leq 37$ for $50$ keV electrons can be found in Fig.\ 4 of \cite{osti_774848} and point towards a cross section for heavy nuclei that are $\sigma_{\rm inel}\simeq 0.5$ \AA$^{2}$ (a crude estimate). It is interesting to consider the possibility of using metalic foils of e.g.\ $^{151}$Sm. Taking the density of Sm, $\rho = 7.5$ g/cm$^3$, we find a number density of $n=3\times 10^{22}$ cm$^{-3}$. For a fixed area $A$ the signal will scale as $S\sim A\times H \times n \times (1- \langle P_{\rm inel}\rangle)$ where $H$ is the height (or thickness) of the foil, and $\langle P_{\rm inel}\rangle$ is the average probability of scattering inelastically while exiting the foil. As a rough proxy we may take $\langle P_{\rm inel}\rangle = \exp(- H n \times (0.5) \text{\AA}^2)$. The optimal foil thickness to maximize the signal is then given by $H_{\rm opt} = n \times (0.5) \text{\AA}^2 \approx 7$ nm; this corresponds to roughly $30$ $^{151}$Sm atoms.  A 1 cm$^2 \times$ 7 nm foil of $^{151}$Sm would contain $\sim 2 \times 10^{17}$ Samarium atoms. One would then need $10^{11}-10^{12}$ such foils to fulfil the 10-event per year criterion outlined above. A similar analysis for $^{171}$Tm suggests, that a $6$ nm foil thickness is optimal and that one would need roughly the same number of foils. Given the difficulties in purifying $^{171}$Tm this suggests, at least in our naive implementation, that $^{151}$Sm is the preferable candidate.\newline
Commercially available samarium films are sold as small as $75$ microns, which is four orders of magnitude thicker than the naive estimate above. We note, however, that nanofoils have been successfully produced in recent years with the thinnest gold foils achieving two-atom thickness \cite{nanofoil_gold}, and 10 nm foils being readily available across a range of metals \cite{Li2006,Qu2006,Zhao2018,Zhao2015}. Given the relative ease with which $^{151}$Sm can be produced such thin foils may not be necessary.  Provided one is not limited by raw material, thicker foils are perfectly acceptable because the outer $\sim 10$ nm of the foil (on both sides) can serve as a C$\nu$B target with the inner bulk of a foil serving as an effective substrate. Such a design would be much heavier than the 6 tonne estimate from above, but this extra detector mass would not affect the packing efficiency since the additional $^{151}$Sm would not increase interfoil spacing. We leave design optimization to the experimental collaboration, but conclude that in addition to any intrinsic benefits of heavy nuclei, their chemical and material properties may also offer useful alternatives to the nominal tritium-graphene design being pursued by PTOLEMY.

\subsection{Atomic excitation effects \label{subsec:shake}}

The above event estimates neglect atomic shake-up and shake-off to which we now turn our attention. In the previous section we have focused on the neutrino capture cross section at threshold for an isolated nucleus in free space, $(\sigma v)_\nu$. This neglects all of the atomic dynamics. In reality, a neutrino capture results in a sudden change in the nuclear charge $Z\rightarrow Z+1$, and causes the nucleus to recoil. Combined these result in either shake-up, where an inner atomic electron is excited to a higher level, or shake-off where an outer-shell atomic electron is ionized. Both cases negligibly impact the extraction of the relevant matrix element from the beta decay half-life. By unitarity, these effects just shuffle strength of the decay spectrum to different energies, and this does not meaningfully impact the integral in \cref{eq:vv}.

These effects are extremely important for cosmic neutrino detection. Atomic excitations are $\sim {\rm eV} $ in energy, and so a neutrino capture event that ionizes an outer shell electron will fall below the endpoint of the beta decay spectrum and be invisible. The relevant cross section is therefore $(\sigma_ v)_{\nu-{\rm det}}$ where detection requires that the signal electron has an energy above the endpoint of the beta spectrum. Given a probability of shake-up $P_{\rm SU}$ and a probability of shake-off $P_{\rm SO}$ we have 
\begin{equation}
    (\sigma v)_{\nu-{\rm det}} = (1- P_{\rm SU}-P_{\rm SO}) \times (\sigma v)_{\nu} ~.
\end{equation}

For tritium, both shake-up and shake-off effects are calculable using simple hydrogenic wavefunctions and can be computed using the sudden approximation by calculating the overlap between hydrogenic wavefunctions with $Z=1$ and $Z=2$. The shake-up probability for tritium is around $25\%$ \cite{PhysRevA.48.268,Hayen:2017pwg} whereas the shake-off process for tritium is subdominant and can be neglected.

For heavy nuclei, shake-off dominates over shake-up \cite{PhysRevA.45.6282,Hayen:2017pwg} because inner shell excitations are effectively Pauli blocked, while outer shell orbitals have small binding energies and are more easily ionized. A precise calculation is much more challenging than for tritium because of the many-body nature of a $Z\sim 60$ atom. Modern numerical calculations of atomic wavefunctions can in principle be used, and can obtain sub-percent level precision in some cases (see e.g.\ \cite{Ruiz:shakeoff}),  however this is beyond the scope of our present focus. Nevertheless, it is uncontroversial that the shake-off probability is roughly $25-30\%$ in heavy nuclei \cite{PhysRev.169.27}. For a related discussion see \cite{Nussinov:2021zrj}. 

The typical C$\nu$B discovery estimates are such that 100 grams of tritium are required if PTOLEMY-like experiment runtime is a single year. These estimates rely only on the neutrino capture cross section at threshold and do not account for the loss of C$\nu$B signal electrons below the beta decay endpoint due to atomic excitation. In light of the significant shake-up for tritium we therefore estimate that target mass requirements should be  enhanced roughly by a factor of $4/3$ and the similar quantitative statement holds for $^{171}$Tm and $^{151}$Sm
in the context of shake-off. Hence, while this may impact target designs, it has little impact on target comparisons since all atoms suffer a roughly 20-30\% loss of signal due to atomic excitations. We also note that molecular effects at the endpoint may further enhance the necessary target mass, but we do not consider those here.

\section{Conclusions and outlook} 
\label{sec:conclusions}

Detecting the C$\nu$B is an old problem, with an old solution, yet its inherent technical challenges have inhibited discovery for half a century. A flexible toolbox will help enable future progress by allowing experimental collaborations to compare costs and benefits of various nuclear targets. This may provide an alternative path to discovery if a tritium target faces insurmountable challenges, or it may enable next generation technology that can overcome the low-statistics barrier that C$\nu$B detectors must confront.
In either case, the ability to use and plan for alternative nuclear targets is a benefit to the experimental community. 

In this work we have extracted the neutrino capture cross section at threshold from the precisely measured half life of both $^{151}$Sm and $^{171}$Tm. Our extraction is primarily limited by the validity of the $\xi$ approximation, which is expected to hold at the percent level for both nuclei; this expectation is supported by empirical evidence, especially for $^{151}$Sm for which a high-statistics measurement has been recently performed \cite{Sm}. We have included all other percent-level corrections to the beta decay spectrum as identified and outlined in Table VII of \cite{Hayen:2017pwg} using \verb|BetaShape| as a convenient tool for implementing the bulk of the corrections. 

We have further considered atomic shake-up and shake-off both for tritium and for the heavy nuclei $^{151}$Sm and $^{171}$Tm. We find comparable losses (where the signal electron is lost beneath the beta background) due to atomic excitations. While we have not supplied a percent level determination of the shake-up or shake-off probabilities for either tritium or heavy nuclei these should be computable with modern atomic physics techniques. They should be revisited with state of the art Hartree-Fock calculations. 

Our cross section extraction allows one to estimate the necessary size of a C$\nu$B detector composed of heavy nuclei. The 6 tonne $^{151}$Sm detector is required to achieve the same yearly event yield as 100 gram tritium one. A 350 kg $^{171}$Tm detector would suffice provided $^{170}$Tm is efficiently removed. \newtext{We note that our methodology can be easily extended to other heavy beta emitters with low $Q$-values. One appealing example is 
$^{210}$Pb which has a $Q$ value of 63.5 keV and $t_{1/2} = 22.3$ yr; interestingly the
beta decay of this element is also first-forbidden non-unique. We have also identified 
$^{228}$Ra with $Q=45.9$ keV and $t_{1/2} = 5.75$ yr as an experimentally viable candidate.} 

We have not considered the chemical properties of $^{151}$Sm or $^{171}$Tm; however, this is an essential consideration for practical purposes. For example, Van der Waals binding is proportional to atomic polarizabilty, and these are an order of magnitude larger for heavy nuclei as compared to hydrogen. Binding effects, recently considered in \cite{Cheipesh:2021fmg,Nussinov:2021zrj},  deserve further scrutiny. These considerations ultimately depend on experimental details such as the choice of binding substrate, and may therefore be hardware dependent. If this is the case they are best considered with input from the PTOLEMY collaboration. 

In summary, we have provided a percent level extraction of the neutrino capture cross section at threshold. Our theoretical uncertainty is dominated by the $\xi$-approximation and can be further scrutinized with shell-model calculations; since nuclei under considerations are 
open-shell heavy deformed one should stress that for capturing such deformation one would need to employ a large single-particle space or an evolved effective Hamiltonian.
Improved Hartree-Fock calculations can supply percent level determinations of the shake-up and shake-off probabilities. There is no theoretical limitation in using heavy nuclei to detect the C$\nu$B and the issue is reduced entirely to practical experimental considerations.

\textbf{Note added:} After completing this work we were informed about recent measurements and improved calculations for $^{151}$Sm. Nuclear matrix elements were computed using a shell-model code (NuShell-X  \cite{NuShell-X}) which were then incorporated into the Behrens-B\"uhring formalism. The corrections to the allowed shape from this shell model calculation lie within our quoted error band.  New data suggests that deviations from the predictions of beta shape in the low-energy  ($T_e<5$ keV)  portion of the spectrum can shift the capture cross section predicted herein by a few-percent \cite{Xavier}. 

\section*{Acknowledgements} 
We benefited greatly from discussions with Chris Tully and thank him for his detailed comments on our manuscript. We also thank Alexey Boyarsky for helpful comments and discussions. We would like to thank Xavier Mougeot for sharing recent results on $^{151}$Sm prior to publication. We also thank Leendert Hayen, Brad Plaster, and Javier Menendez for their helpful comments and feedback on our manuscript. RP thanks the Fermilab theory group for their hospitality and support. This work was supported by the U.S. Department of Energy, Office of Science, Office of High Energy Physics, under Award Number DE-SC0019095. This manuscript has been authored by Fermi Research Alliance, LLC under Contract No. DE-AC02-07CH11359 with the U.S. Department of Energy, Office of Science, Office of High Energy Physics. This work was performed in part at Aspen Center for Physics, which is supported by National Science Foundation grant PHY-1607611.

\bibliography{biblio.bib}

\end{document}